\documentclass[aps, twocolumn,nofootinbib, showpacs]{revtex4-1} 
\usepackage[unicode=true,pdfusetitle,
bookmarks=true,bookmarksnumbered=false,bookmarksopen=false,
breaklinks=false,pdfborder={0 0 0},backref=false,colorlinks=false] {hyperref}
\hypersetup{ colorlinks,linkcolor=myurlcolor,citecolor=myurlcolor,urlcolor=myurlcolor}
\usepackage{braket,colortbl,amsthm,amsmath,cleveref,amssymb,txfonts}\definecolor{myurlcolor}{rgb}{0,0,0.7}
\usepackage{graphics,graphicx}
\usepackage{color}
\usepackage{amssymb}
\usepackage{amsthm}
\usepackage{amsfonts}
\usepackage{float}
\usepackage{graphicx}
\usepackage[format = plain,labelfont = bf,up, textfont = normal , up, justification
=raggedright, singlelinecheck =false]{caption}
\usepackage{subcaption}
\usepackage{tabularx}
\usepackage{amsmath}
\usepackage{braket}

\usepackage{graphicx}
\usepackage[utf8x]{inputenc}
\usepackage{color,soul}
\usepackage{amsmath}
\usepackage{braket}
\usepackage{latexsym}
\usepackage{amssymb}
\usepackage{amsthm}
\usepackage{xcolor}
\usepackage{bm}
\usepackage{graphics,epstopdf}
\usepackage{color}
\usepackage{enumitem}
\usepackage{fmtcount}
\usepackage{booktabs}
\usepackage{epsfig}
\usepackage[normalem]{ulem}
\theoremstyle{plain}

\def\bea{\begin{eqnarray}}
\def\eea{\end{eqnarray}}
\def\ba{\begin{array}}
\def\ea{\end{array}}

\def\ket{\rangle}

\begin{document}
%\title{Quantum information can remain without physical body in a volatile form}

\title{Quantum information can remain without physical body in  volatile form}

\author{Brij Mohan, Sohail, Chirag Srivastava, Arun K. Pati, Ujjwal Sen}

\affiliation{Quantum Information and Computation Group,
Harish-Chandra Research Institute, HBNI, Chhatnag Road, Jhunsi, Allahabad, India}
%In the era of second generation of quantum technology d
\begin{abstract}
A deeply rooted view in classical and quantum information is that 
``information is physical'', i.e., to store and process information, we need a physical body.
Here we ask whether quantum information can remain without a physical body. We answer this question in the affirmative, i.e., we argue that quantum information can exist without
a physical body in  volatile form. We introduce the 
 notion of the volatility of quantum information and show that indeed the conditions for it are 
  naturally satisfied in the  quantum 
 teleportation protocol. 
 We argue that even if special relativity principles are not assumed, it is possible to make quantum information volatile. 
%We discuss another plausible scenario where the volatility  of quantum information can appear. 
We also discuss the classical limit of the phenomenon, as well as the multiparty scenario.

\end{abstract}
\maketitle

\section{Introduction}
Quantum information science
%theory 
aims to understand  novel properties of the quantum world and harness these features in 
%various 
quantum information processing 
%tasks. 
devices.
%Such novel 
These 
properties have no classical counterparts, and 
%these 
are purely non-classical manifestations of quantum physics. In the quantum world, the entity 
``quantum information'' symbolizes - in general understanding - the information content inscribed in a quantum state of a physical system. Thus, information and quantum state are the two primary constituents of 
quantum information. 
%theory. 
%practice and theory.
Unlike the information in a pure classical bit, quantum information in a pure quantum state is not fully accessible.
%, and e.g., a 
A single pure qubit has potentially 
an infinite number of bits of information, but one cannot extract it by using measurements on a single qubit. 
We have learnt to reconcile with the situation that while the description of a single qubit requires an infinite number of bits, only a single bit can ever be extracted from a single copy of a qubit~\cite{mercurousnitrate}
%\textcolor{red}{[pls refer to Refs. 3 and 6 of https://journals.aps.org/pra/pdf/10.1103/PhysRevA.74.052332]} 
- seemingly a form of ``bound'' information,  distinct from the ones in Refs.~\cite{acharyya-prafulla-ray,badalrajani}.
%\textcolor{red}{[pls refer to P. Horodecki, Phys. Lett. A 232, 333
%1997; M. Horodecki, P. Horodecki, and R. Horodecki, Phys. Rev.
%Lett. 80, 5239 1998;
%N. Gisin and S. Wolf, Proceedings of Crypto 2000, Lecture Notes in Computer Science 1800, 482 (2000)]}.
It is also believed that this inaccessibility of the quantum information is 
related to the existence of ``non-local'' correlations \cite{Bell'64}, which do not defy causality \cite{Jozsa'98,nilotpal}. Therefore, understanding various 
limitations on quantum information has been a major challenge for researchers in the last two decades or so \cite{Wootter'82,Dieks'82,sawal-jawab,Pati'00,common-origin,samal,sam,Modi'18,punDarikaksha-purakayastha-1,punDarikaksha-purakayastha-2,punDarikaksha-purakayastha-3}. 

%Ben-Shor'98

 Whether we speak of classical 
 %information 
 or 
 quantum information, after Landauer's famous insight that ``information is physical'', we inevitably think that information has to 
 %remain 
 reside
 in a physical body. If the physical body obeys classical mechanics, we say that we are dealing with classical information, and if the physical body is governed by quantum mechanics, then we state that we are 
 handling
 %treating 
 %deal with 
 quantum information.\footnote{We note that a quantum system can encode a bit or a qubit, and so one often speaks of classical information encoded in a quantum system, and which provides interesting nonclassical applications like the dense coding protocol~\cite{mrinal1,mrinal2}.} Landauer’s view was that when we 
talk about information, we need to inscribe it somehow in a physical body. This view has been firmly held in the physics community, especially after 
%Landauer 
%proved 
it was understood that %the relation
%between information erasure and 
a minimum amount of work is needed to erase one bit of information, and it equals 
%, i.e., to erase a single bit we need to spend 
%$  K_BT \ln 2$ 
\(k_BT\)
amount of 
%entropy,
energy,
where $k_B$ is the Boltzmann constant and $T$ is 
the related absolute temperature~\cite{Landauer1961,Landauer1991,Landauer1996,Landauer1999}.
%. Thus, the erasure of one bit of information generates heat, and increases physical entropy which is the famous Landauer’s Principle 
%\cite{Landauer1961,Landauer1991,Landauer1996,Landauer1999}. 
See also~\cite{akash-bhara}.

Quantum teleportation \cite{Bennett'93} is a prime quantum information processing protocol, where various counter-intuitive aspects of quantum information 
can be 
and has been 
discussed and 
debated. In the standard quantum teleportation protocol, information about an unknown qubit can be sent from one location to another by using a maximally entangled pair \cite{EPR} and 
two bits of classical 
communication. Even though an unknown qubit 
%contains
requires 
an infinite number of bits of classical information to be described,  one can transfer the whole 
information by sending just two classical bits, provided the pre-shared entangled state~\cite{Horo'09,Guhne'09,Das'17} is available. In the literature, various arguments have been provided on how the quantum information encoded in a qubit can be transferred by communicating merely 
two bits of classical information from the sender's to the receiver's location. In Ref.~\cite{Jozsa'98}, it is interpreted that the quantum information flows across the 
entanglement  which is destroyed in the process. In Ref.~\cite{Penrose'98}, an idea of quantum information traveling backward in time is 
presented in order to justify the quantum teleportation. There also exist claims on refuting any non-local influences causing information flow in quantum teleportation 
\cite{Hayden'00}.

Even though it is a rather deep-rooted view that both classical 
%information 
and 
quantum information %have been deeply rooted with the view that information is 
are
physical, we ask whether 
quantum information can remain without a physical body?  We answer this question in the affirmative, i.e., we argue that quantum information can exist without
a physical body, in  volatile form. We introduce the 
 notion of the volatility of quantum information, and show that the conditions for volatility of quantum information is satisfied in the quantum 
 teleportation protocol~\cite{Bennett'93}. 
 %
 %
% \textcolor{red}{[this part shud probably go!] We should mention that our argument does not necessarily %vindicate 
% dispute
% Landauer's view. Rather, it opens up a new possibility in addition to what is implied by the 
% Landauer's important insight, viz., quantum information has a dualistic manifestation  - sometimes it needs a physical body and sometimes not, with the former case - ``Landauer case'' - holding under certain circumstances.} 
 %
 %
We argue that the conditions for volatility hold irrespective of whether we assume the laws of special relativity. We show that volatility can occur also for classical systems, although the situation is fundamentally different with the quantum case. In both quantum and classical scenarios, volatility is shown to  occur in bipartite as well as in multiparty situations, with the no-cloning~\cite{Wootter'82, Dieks'82, sawal-jawab} and no-broadcasting~\cite{punDarikaksha-purakayastha-1} theorems making the multiparty cases have an essential difference with the bipartite ones. Although the specific cases of volatility described in the paper are related to teleportation protocols, we provide a certain plausible situation beyond the teleportation protocol where volatility may occur.
%\textcolor{cyan}{[will check this part later] Towards end, we argue that even if special relativity principles are not assumed, it is possible to make quantum information volatile. 
%We discuss several other scenarios where the volatile nature of quantum information can happen.}
%
%
We believe that this 
may
have 
%far reaching 
important
consequences on the very 
notion of ``information'' in quantum theory 
and beyond.

%\textcolor{cyan}{[will check this later!] 
The rest of the paper is organized as follows.  In Sec~\ref{sec2}, we formalize the notion of volatility of quantum information. 
%, i.e., 
%whether it is possible to have a scenario where information can remain without the physical system. We argue that it is indeed possible and
%then unveil that it originates naturally in the quantum teleportation protocol. We also argue that even if principles of 
%special relativity are not assumed, then also quantum information can remain in volatile form.
%In Sec.\textcolor{red}{???}, %\ref{sec3}, 
%we provide other scenarios where quantum information exists in volatile form.
In Sec.~\ref{tumi-sandhyar-meghmala}, we show how volatility can appear in the classical case.
The multiparty case is considered in Sec.~\ref{lal-neel-holud}.
%multiparty case.
%Sec.~\ref{dilip-ghosh} beyond tele.
We discuss our results  in Sec.~\ref{sec3}, where we also provide an example beyond the teleportation scenario where volatility may plausibly appear.
%we conclude.}

\section{Volatility of quantum information}  \label{sec2}

Volatility of quantum information is attained through a quantum channel, belonging to what we call the family of volatility-rendering quantum channels (VQCs), that allows us to render the information in a quantum state into a form that 
%that quantum information 
is neither in any of the subsystems of a system nor in the 
correlations between them. It is to be noted that this scenario is in stark contrast
to both hiding~\cite{sam} and masking~\cite{Modi'18} of quantum information. In the hiding map, information disappears from one subsystem and 
one asks the question whether information can remain in the quantum correlation or in the other subsystem. In  masking, we encode quantum information in a bipartite entangled 
state and ask the question whether we can keep quantum information only in correlation and not in either subsystem. The notion of volatility of quantum information goes 
one step further.
%, and it is quite amazing to see that quantum information can remain indeed in the strange form.

Consider a physical system in control of an observer, Alice, with the Hilbert space associated with the system being \(\mathcal{H}_A\). Another observer, called Bob, is in possession of his own physical system,
%. For ease of notation and discussion, we will pretend that there is a single Bob, 
with the associated Hilbert space being \(\mathcal{H}_B\). Alice and Bob
% two parties Alice and Bob with their associated Hilbert space $\mathcal{H}_A$ and $\mathcal{H}_B$ respectively. Let they 
share an entangled state $\rho_{AB}$, which 
acts on the Hilbert space $\mathcal{H}_A \otimes \mathcal{H}_B$. We now consider yet another physical system, with the associated Hilbert space being \(\mathcal{H}_a\), and which is in the state \(|\psi\rangle_a\). This physical system is again in control of Alice, and this is the state whose information we wish to make volatile.
%Let $|\psi \rangle_{a} \in \mathcal{H}_{a}$ is the input state which we need to make volatile. 
There is a classical communication channel between Alice and Bob, and they are allowed arbitrary quantum mechanical operations in their respective laboratories, i.e., on physical systems corresponding to the Hilbert spaces \(\mathcal{H}_a \otimes \mathcal{H}_A\) and \(\mathcal{H}_B\) respectively, or on their local extensions. This is what is refereed to as local quantum  operations and classical communication (LOCC) between Alice and Bob.

Making quantum information volatile is an LOCC-based quantum map,
%local quantum operation and classical communication (LQOCC), where these conditions hold:
\begin{equation}\label{yaad}
|\psi \rangle _{a} \langle \psi | \otimes \rho_{AB} \rightarrow T_t(|\psi \rangle_{a} \langle \psi | \otimes \rho_{AB}),
\end{equation}
where $T_t$ is an LOCC (in the Alice to Bob partition) acting for the time duration $t_{1} \leq t \leq t_{2}$. Otherwise, $T_{t}=\mathbb{I}_{aAB}$, the identity superoperator on the entire system. 
We intend to make the quantum information in 
%Therefore, we say that the quantum information 
$|\psi \rangle _{a}$ associated with $\mathcal{H}_a$  into a volatile form for the time duration $t_{1} < t < t_{2}$. We claim that such volatility of the quantum information in \(|\psi\rangle_a\) is achieved if  
%when 
\begin{itemize}
\item[(i)] the reduced states, 
 $\rho_{aA}=\text{Tr}_{B}  [T_t(|\psi \rangle_{a} \langle \psi | \otimes \rho_{AB})]$ 
 and 
% $\rho_{A}=\text{Tr}_{aB}  [T_t(|\psi \rangle_{a} \langle \psi | \otimes \rho_{AB})]$ and 
$\rho_{B}=\text{Tr}_{aA}  [T_t(|\psi \rangle_{a} \langle \psi | \otimes \rho_{AB})]$, are independent of $|\psi \rangle_a \langle \psi|$, for \(t_1 < t < t_2\),
\item[(ii)] any correlations that may remain between \(aA\) and \(B\) cannot account for the quantum information in \(|\psi\rangle_a\), and 
\item[(iii)] the state \(|\psi\rangle\) can be retrieved somewhere after \(t=t_2\).
\end{itemize}
%\noindent 
\emph{Remark.} First of all, a quantification will be needed for item (ii), which will be done later. Secondly, the above formulation in item (i) is qualitative, and in principle, can be quantified, although we do not walk that path. Thirdly, item (iii) is needed to avoid the trivial VQC  where one performs a projection-valued  measurement on the system corresponding to \(\mathcal{H}_a\) at \(t=t_1\) onto a standard basis, independent of \(|\psi\rangle\).\\\\
These conditions will imply that the quantum information was present in a part of the system for \(t\leq t_1\) and is there also for \(t \geq t_2\). But for \(t_1 < t < t_2\), the same is not present in in $aA$
%, `$A$' and 
or $B$, and any correlations between \(aA\) and \(B\) is not enough to account for it. In such circumstances, we say that the quantum information is in volatile form duirng \(t_1 < t < t_2\).
%At $t=t_{1}$, local operation (LO) may destroy entanglement between A and B. Hence, quantum information is not in the correlation.
%If we can ensure these conditions then we will say that quantum information can exist in a volatile form.

%\textbf{\emph{Teleportation Protocol:}}

We claim that volatility is   indeed seen in the quantum teleportation protocol~\cite{Bennett'93}, which we briefly describe here, for completeness and also for later reference in the paper. 
Suppose that the two distant observers, Alice and Bob,
%Now, we show that indeed there is a scenario where the above conditions hold true. 
%Consider the standard quantum teleportation protocol proposed by C. H. Bennett \emph{et. al.} \cite{Bennett'93}, where two distant observers, Alice and Bob, 
share the maximally 
entangled state,
\begin{equation}
|\Phi^{+}\ket_{AB}=\frac{1}{\sqrt{2}}(|00\ket+|11\ket),
\end{equation}
 where $|0\ket$ and $|1\ket$ are orthonormal quantum states.
% Henceforth, we assume that there is exactly a single Bob, unless  explicitly mentioned
% otherwise. 
The task is to transfer the information in an arbitrary pure qubit, 
 $|\psi\ket_{a}=\cos\frac{\theta}{2}|0\ket+e^{i\phi}\sin\frac{\theta}{2}|1\ket$, from Alice's lab to Bob's lab, where $\theta \in [0,~\pi]$ and $\phi \in [0,~2\pi)$, without actually transporting any quantum system between the labs, although a classical communication channel from Alice to Bob can be used for a finite (preferably, low) number of rounds. Only a single copy of the state \(|\psi\rangle\) is provided to Alice. It is to be noted that if the state \(|\psi\rangle\) is unknown to Alice, she cannot clone it, and so, cannot estimate its identity via quantum measurements. See~\cite{titumir,majdoor} in this respect. Note also that even if Alice is able to know the identity of the state, she would need an infinite amount of classical communication to send that identity to Bob using the classical channel only, i.e., without using the shared entangled state. 
 To proceed, Alice performs a measurement in the Bell basis, $\{|\Psi^\pm\ket_{aA},~|\Phi^\pm\ket_{aA}\}$, on the physical systems corresponding to which the Hilbert space is \(\mathcal{H}_a \otimes \mathcal{H}_A\),
 %measurement on the qubit 
 %$|\psi\ket_A$ and her part of the shared entangled state, $|\Phi^{+}\ket_{AB}$, 
 where 
\begin{eqnarray}
|\Phi^\pm\ket_{A'A}=\frac{1}{\sqrt{2}}\left(|00\ket \pm |11\ket\right), \nonumber \\
|\Psi^\pm\ket_{A'A}=\frac{1}{\sqrt{2}}\left(|01\ket \pm |10\ket\right).
\end{eqnarray}
%
%\begin{table}[h!]
%
%  \begin{center}
 %   \caption{The standard quantum  teleportation protocol. 
  %  Here, \(I_2\) represents the identity operator on the qubit Hilbert space, while \(\sigma_i\) are the Pauli spin matrices. The outcomes occur with equal probabilities.
    %Application of certain unitary by Bob depending on Alice's outcome.
   % }
    %\label{tab1}
    %\begin{tabular}{|c|c|c|c|}
    %\hline 
     % i& \textbf{Alice's outcome} & \textbf{Bob's state} & \textbf{Bob's unitary} \\
      %
      %\hline\hline
%0&
%$|\Phi^+\ket_{aA}$ & $|\psi\ket_B$ & $I_2$\\[0.8ex]
% \hline
%1&
%$|\Phi^-\ket_{aA}$ & $\sigma_z|\psi\ket_B$ & $\sigma_z$\\[0.8ex]
% \hline
%2&
%$|\Psi^+\ket_{aA}$ & $\sigma_x|\psi\ket_B$ & $\sigma_x$\\[0.8ex]
% \hline
% 3&
% $|\Psi^-\ket_{aA}$ & $\sigma_y|\psi\ket_B$ & $\sigma_y$\\[0.8ex]
% \hline
%    \end{tabular}
%  \end{center}
%\end{table}
%
After the Bell state measurement, Alice sends the information about her measurement outcome (two bits of classical information) to Bob. 
Depending on Alice's outcome, Bob applies a certain unitary. 
Let the classical information be denoted by \(i\), which takes the values \(0\), \(1\), \(2\), \(3\), corresponding, respectively, to the outcomes \(|\Phi^+\rangle\), \(|\Phi^-\rangle\), 
\(|\Psi^+\rangle\), \(|\Psi^-\rangle\), in the Bell measurement. 
%as presented in Table~\ref{tab1}, 
To 
%retrieve 
resurrect 
the information in the quantum state \(|\psi\rangle\) in his own lab, and which was initially present in Alice's lab, Bob does nothing for \(i=0\), while he applies \(\sigma_z\), \(\sigma_x\), \(\sigma_y\), the Pauli matrices, for \(i=1, 2, 3\) respectively.  
%After this measurement, the particle at Bob's possession has a state out of the states, $|\psi\ket_B,~\sigma_x|\psi\ket_B,~\sigma_y|\psi\ket_B,$ and $\sigma_z|\psi\ket_B$, 
%Depending on Alice's outcome $|\Phi^+\ket_{A'A},~|\Psi^+\ket_{A'A},~|\Psi^-\ket_{A'A},$ and $|\Phi^-\ket_{A'A}$, . Therefore, Bob waits for the two bits classical information by Alice about the measurement outcome, and then performs a correcting unitary to get back the quantum state $|\psi\ket_{B}$ at his laboratory. For example, if Alice has an outcome $|\Phi^-\ket_{A'A}$, then after the classical communication, Bob performs $\sigma_z$ on his particle to retrieve $|\psi\ket_B$. \\
Note that the outcomes of the Bell measurement occur with equal probabilities. We assume that the Bell measurement is performed by Alice at time \(t=t_1\), and she sends the information about its outcome to Bob at some time in the interval \([t_1, t_2)\). The information reaches Bob at \(t=t_2\), and he immediately applies the requisite unitary according to the rule described earlier. 
%Table~\ref{tab1}. 
We assume that the protocol, and in particular the time at which  the measurement is performed and the time at which the unitaries are applied, are known beforehand to all, including to the two observers.

Going back to the notations related to Eq.~(\ref{yaad}), in case of quantum teleportation, \(\rho_{AB} = |\Phi^+\rangle \langle \Phi^+|\), and \(T_t\) represents the LOCC actions in the teleportation protocol. Therefore, we have 
\begin{equation}\label{jhara-mala}
|\psi \rangle _{a} \langle \psi | \otimes |\Phi^+\rangle_{AB} \langle \Phi^+| \rightarrow T_t(|\psi \rangle_{a} \langle \psi | \otimes |\Phi^+\rangle_{AB} \langle \Phi^+|),
\end{equation}
where \(\mbox{tr}_{AB} T_t(|\psi \rangle_{a} \langle \psi | \otimes |\Phi^+\rangle_{AB} \langle \Phi^+|) = |\psi\rangle_a \langle \psi|\) for \(t< t_1\), and \(\mbox{tr}_{aA} T_t(|\psi \rangle_{a} \langle \psi | \otimes |\Phi^+\rangle_{AB} \langle \Phi^+|) = |\psi\rangle_B \langle \psi|\) for \(t\geq t_2\).
We can therefore say that before Alice does the Bell measurement, $|{\psi}\ket $ was in Alice's lab, although Alice may or may 
not know what it is.
%, but the full information %(whatever that may mean, as we still do
%not know how to quantify quantum information of a pure (or mixed) quantum state) 
%about $|{\psi}\ket $
%was at Alice's location. 
After Alice does the Bell measurement, at  time $t=t_1$, and before the classical communication 
reaches Bob's lab, at  time $t = t_2$, 
%we can say that 
$|{\psi}\ket $ is neither at Alice  nor at Bob. This is because Alice's two
qubits are in one of the Bell states and hence does not have any
information about $|{\psi}\ket $, and Bob's qubit is still in an equal mixture of the state \(|\psi\rangle\) and the three Pauli-rotated \(|\psi\rangle\)s 
%(see Table~\ref{tab1}) 
which equals $\frac{1}{2}I_2$ and has no information about
$|{\psi}\ket $. Here, \(I_{2}\) is the identity operator on the qubit space. 

%\begin{equation}
%Cu_2SO_4 \to_{\Delta} \chi + \phi
%\end{equation}
%Snajh-ke latex shekhachchhilum

% \textcolor{red}{(para can be removed)    Then, the question is, where is $|{\psi}\ket $?  
% We will argue that for that much time (i.e., at least the time it takes CC to reach from Alice to Bob), quantum information may remain in `volatile' form, i.e., it can 
% remain without the physical body.
%One can see that in the teleportation protocol the LOCC operation indeed satisfies the properties that $T_t$ satisfies, given in \eqref{yaad}. Therefore, for time $t_{1} \leq t < t_{2}$, quantum information is indeed in a volatile form.
%At $t_{1}$, Alice does LO which is Bell measurement . At $t_{2}$, CC reaches Bob. So $t_{2}-t_{1}$ is the time CC takes from Alice to Bob.
%In teleportation, $\rho_{AB}=|\Phi^{+} \rangle \langle  \Phi^{+}|$ (say) and we know that quantum teleportation is a scheme where LOCC can be used to describe the process.
%}

The information about a system can of course stay in the correlations between the parts of the system, both for classical and quantum systems. Consider a charged  conducting material,  with the negativity or positivity of the charge being as yet unknown. Let us name this part as \(\alpha\). We bring that material   in contact with another neutral 
conducting material, which we name as \(\beta\). We still do not know the sign of the charge, and the entire two-part system can be represented as \begin{equation}
\label{alakananda}
\rho^{cl}_{\alpha \beta} = \frac{1}{2}(|--\rangle\langle --| + |++\rangle\langle ++|),
\end{equation}
where we are concerned only with the signs of the charges of the two parts.
For this state, the local states are completely depolarized, and devoid of any information. However, the information about what is the sign of the charge, which is a single bit of information, is present in the classical correlation between the two parts of the system, and it can again be quantified, in several ways, as one bit. One way to quantify the amount of classical correlation in a state is to consider the relative entropy ``distance''~\cite{mono-dilo-na-bnodhu} between the state and the nearest product state~\cite{mono-je-nilo-shudu, jadi-amai-nai}.

Therefore, we now need to look at the correlations that may exist between \(aA\) and \(B\) during \((t_1, t_2)\). The state of the entire system during this time is 
\begin{equation}
\label{kothai-kothai-je-raat-haye-jai}
\frac{1}{4}\sum_{i=0}^3 |\Psi_i\rangle_{aA}\langle\Psi_i| \otimes |\psi_i\rangle_{B}\langle\psi_i|,
\end{equation}
where 
\(|\Psi_0\rangle = |\Phi^+\rangle\),
\(|\Psi_1\rangle = |\Phi^-\rangle\),
\(|\Psi_2\rangle = |\Psi^+\rangle\),
\(|\Psi_3\rangle = |\Psi^-\rangle\), and 
\(|\psi_0\rangle = |\psi\rangle\),
\(|\psi_1\rangle = \sigma_z|\psi\rangle\),
\(|\psi_2\rangle = \sigma_x|\psi\rangle\),
\(|\psi_3\rangle = \sigma_y|\psi\rangle\).
%
%
%the states can be read off from Table~\ref{tab1} by noting the value of \(i\). 
This is the state of the entire system for Bob before the classical communication reaches him, and for anyone else who hasn't yet received the same. The state of the entire system for Alice is of course a pure one, with its identity being dictated by the  outcome of the Bell measurement.
The state in Eq.~(\ref{kothai-kothai-je-raat-haye-jai}) is a ``classical-quantum'' state. It is not entangled~\cite{Horo'09,Guhne'09,Das'17}. Among the quantum correlations beyond entanglement~\cite{nonigopal}, quantum discord~\cite{jadi-amai-nai} and one-way quantum work deficit~\cite{bidesh-bose} are also vanishing, if the optimized measurements corresponding to these definitions are carried out on Alice's side. However, quantum discord and one-way quantum work deficit
%other measures of quantum correlations 
can be nonzero if measurements are carried out on Bob's side, although this is somewhat artificial, as the classical communication channel needed to check for such quantum correlation runs from Bob to Alice, which is opposite to the direction of the classical communication channel in the standard quantum teleportation protocol. 
%in particular for the smeasurements on the other side, and for the two-way quantum discord
We therefore regard this state as without any quantum correlation, that is relevant to the purpose at hand. 

The state of the entire three-qubit system of Alice and Bob during \((t_1,t_2)\), is certainly classically correlated. It is exactly this classical correlation that makes the classical communication from Alice to Bob of  relevance for the teleportation protocol. However, this classical correlation is bounded by the dimension, in bits, of the smaller part of the shared quantum system, which is that of Bob. Bob has a two-dimensional quantum system, and therefore can support only two orthogonal states, so that the classical correlation, according to several definitions but in particular to the relative entropy -based one described earlier, is bounded by a single bit. 
It is intriguing that even though the shared state after the Bell measurement has a single bit of classical correlation, the quantum teleportation of a qubit requires two bits to be transferred for the success of the protocol.
This single bit cannot hold in it the information about the two real numbers \(\theta \in [0,\pi]\) and \(\phi \in [0,2\pi)\).

We therefore claim that the conditions in items (i), (ii), and (iii) required for volatility of quantum information are all met during the time \((t_1,t_2)\) of the standard quantum teleportation protocol.
%, with the VQC being 
%
%In quantum teleportation, we have $\rho_{AB} = | \Phi^+ \rangle _{AB} \langle \Phi^+ |$ and it is a scheme where LQOCC can be used to describe the process, i.e., 
%\begin{equation}\label{yaad}
%|\psi \rangle _{a} \langle \psi | \otimes | \Phi^+ \rangle _{AB} \langle \Phi^+ |  \rightarrow T_t(|\psi \rangle_{a} \langle \psi | \otimes | \Phi^+ \rangle _{AB} \langle \Phi^+ | )
%\end{equation}
%with $\Tr_{aA} [T_{t_2} (|\psi \rangle_{a} \langle \psi | \otimes | \Phi^+ \rangle _{AB} \langle \Phi^+ | ] =  |\psi \rangle _{B} \langle \psi | $ .

One may argue that after the Bell state measurement, since Alice knows the measurement outcome, for her, $|\psi\rangle$ (up to a local unitary) is already in Bob's lab.
But, quantum state is an observer-dependent notion. Therefore,
even if Alice knows that at Bob's place, the state is $|\psi\rangle$ (up to a local
unitary), that is useless for all practical purposes. As far as \emph{local}
observers are concerned, for the time $t_1 < t < t_2$,
the information about $|\psi\rangle$ is neither in Alice's lab nor in Bob's. 
%It is indeed in a volatile form, i.e., quantum information exists without physical body.
%I think, we need to stress that description of a quantum state is local and observer dependent. 
%Like what Alice describes about her particle which she can manipulate. Remote description of quantum state is
%meaningless as she cannot do anything with that, because she does not have access to Bob's qubit.

It is to be noted that if we believe in special relativity, then the time-interval \(t_2-t_1\) is a nonzero number  bounded below by the quotient of the distance between Alice's and Bob's labs, and the velocity of light. 
%
%From the arguments made above and the principles of special relativity, we conclude that there exists a minimum time interval (the time light takes to travel from Alice's 
%lab to Bob's lab) when the quantum information about $|\psi\ket$ is neither in Alice's lab nor in Bob's lab, it is in the volatile form. 
Even if we do not assume special 
relativity, Alice still has the choice of not sending the two bits of classical information to Bob for a finite period of time. 
%So, in this scenario again, the quantum 
%information about $|\psi\ket$ over this time period is neither in Alice's lab nor in Bob's lab; it is in some volatile form. 
Note also that this information in volatile form during \((t_1,t_2)\) cannot  
be retrieved by any other observer in the universe except Bob, by using the two classical bits sent by Alice. So the information in the volatile form, in the case of the standard teleportation protocol, is somehow 
related to the particle at Bob's lab. It means if the two classical bits are sent to any other observer (say Charu), then she will not be able to resurrect 
%get the full 
%information about 
the unknown quantum state $|\psi\ket$, although she will come to know of the state at Bob's lab. 
%This two classical bit is useless for Charlie.

%(ARUN: THIS PART BELOW IS NOT CLEAR TO ME)

It may be interesting here to mention the following fact about extraction of work from a quantum state~\cite{Landauer1961} (cf.~\cite{bidesh-bose}). Suppose that an observer is provided a pure state of a qubit, but is not told the identity of the state. In such a situation, she will not be able to extract work from the state, where extractable work can be conceptualized by using ``negentropy'', i.e., the von Neumann entropy of the quantum state subtracted from the dimension of the system in bits. The von Neumann entropy of a quantum state \(\varrho\) is given by \(-\mbox{tr}(\varrho \log_2\varrho)\). The above definition of work can be understood by the following argument. For a pure qubit in the state \(|\psi\rangle\), we can assume that it represents the occurrence of a particle in the left half of a box. The information that the particle is in the left half, can be used to move a barrier between the two halves to the right end of the box, which can in turn be used to pull up a weight. At the end of this process, the particle can be either in the left or  the right half of the box, and therefore the corresponding state has maximal entropy among qubit states, while the initial state had vanishing entropy.
The relation of work extraction with volatility is that if Alice knows \(|\psi\rangle\), sends (teleports or transports) it to Bob, but doesn't tell what \(|\psi\rangle\) is, Bob will not be able to extract any work by using the sent state. So, this is similar to volatility (from Bob's perspective), but different, because 
%of what u say below. maybe we shud say this explicitly 
%
in the standard teleportation protocol, the state \(|\psi\rangle\) remains unknown to Bob even after the completion of the protocol, irrespective of whether Alice knows it, so that even after resurrecting the state \(|\psi\rangle\), Bob will not be able to extract any work.

% Similarly, one can think of another situation if Alice prepares a non-passive quantum state $|\phi\rangle$ and send it to Bob. According to the Landauer's principle,
% Bob can
% not extract any work from a given quantum state $|\phi\rangle$ unless he gets the information about the quantum state \cite{Landauer1961}. 

\section{The classical case}
\label{tumi-sandhyar-meghmala}

It is important to mention that volatility of information occurs also in the classical world, but with a difference. 
Indeed, consider all probabilistic mixtures of the orthonormal states \(|0\rangle\) and \(|1\rangle\), which could represent, e.g., the negative and positive charges considered in the preceding section. An arbitrary such mixture can be represented as \begin{equation}
\sigma_{\tilde{a}}=p|0\rangle\langle 0| + (1-p)|1\rangle\langle 1|,
\end{equation}
where \(p \in [0,1]\). They form, for varying \(p\), a set of commuting states, which however are not orthogonal. They cannot be cloned, although can be broadcast~\cite{punDarikaksha-purakayastha-1}. See Ref.~\cite{majdoor} in this respect.
The suffix \(\tilde{a}\) of
\(\sigma_{\tilde{a}}\) indicates the 
%be the state of 
name of the system. This state can be teleported from one location (\(\alpha\)) to another (\(\beta\)) by using the classically correlated shared state
\begin{equation}
\delta^{cl}_{\alpha \beta} = \frac{1}{2}(|00\rangle\langle 00| + |11\rangle\langle 11|),
\end{equation}
which is the same as the state \(\rho^{cl}_{\alpha \beta}\) considered earlier in Eq.~(\ref{alakananda}). A classical communication channel is also needed between \(\tilde{a}\alpha\) and \(\beta\). This protocol 
%may already have been known, but 
is given explicitly in Ref.~\cite{shabnam} (see also~\cite{coronavirus} in this regard). 

In this case again, the real number \(p \in [0,1]\) can remain volatile in the duration between a measurement at \(\tilde{a}\alpha\) and the information about its outcome reaching \(\beta\). The difference with the quantum case is that in the classical case, one needs to go over to mixed states (and so, systems about which we have lower than complete information that is possible to obtain about it) to obtain the phenomenon of volatility. The standard teleportation protocol can of course teleport any mixed state of the qubit system \(a\) by using the same protocol as described before. Such a mixed state can, e.g., be expressed as \(p|\psi\rangle\langle \psi| + (1-p)\frac{1}{2}I_2\), where \(p\in [0,1]\). And in parallel, the same way that made the quantum information in \(|\psi\rangle\) volatile, can additionally make the parameter \(p\) also volatile.

\section{The multiparty case}
\label{lal-neel-holud}

The considerations in the preceding two sections can be carried over to the multiparty case, i.e., a single sender and multiple receivers. 
The volatility-rendering quantum channel in Eq.~(\ref{yaad}), and the items (i)-(iii) required for volatility of quantum information holds also in the multi-receiver case, with the additional restriction that the receivers (along with the sender) are allowed to act locally and communicate classically between themselves.

In the quantum case, for teleportation to work, one needs to consider the genuinely multiparty entangled Greenberger-Horne-Zeilinger state~\cite{asamayer-kuhu},
\begin{equation}
\frac{1}{\sqrt{2}}(|00 \ldots 0\rangle + |11 \ldots 1\rangle),
\end{equation}
as the shared state between the sender and the receivers. One also needs classical communication channels from  the sender to all the receivers. In the classical case, the shared state needed is 
\begin{equation}
 \frac{1}{2}(|00 \ldots 0\rangle\langle 00 \ldots 0| + |11 \ldots 1\rangle\langle 11\ldots 1|).
\end{equation}
However,  at the end of the teleportation protocol,  the state of the receivers is not a product of copies of the state that we obtained as the outcome in the case of a teleportation with a single receiver, neither in the quantum nor in the classical case. In the quantum case, this is blocked by the no-cloning theorem, and the state \(|\psi\rangle\) that was initially at the sender will now be shared as an entangled state between all the receivers. See~\cite{hridayharan,kailash} in this regard.
In the classical case, the states, \(\sigma_{\tilde{a}}\), cannot be cloned, but can be broadcast, and so we get the state 
\begin{equation}
p|00 \ldots 0\rangle\langle 00 \ldots 0| + (1-p)|11 \ldots 1\rangle\langle 11\ldots 1|), 
\end{equation}
after the teleportation protocol
and not 
\(\sigma_{\beta_1} \otimes \sigma_{\beta_2} \otimes \ldots \), 
where \(\beta_1\), \(\beta_2\), ... are the names of the receivers in the classical case. See Ref.~\cite{majdoor}.
The volatility of  information in the multiparty case remains the same as in the single receiver case, except that when the classical communication finally reaches the receivers, the information that was volatile before gets resurrected among several receivers in a shared mode.

%(Do we want to discuss multiparty case?) classical different from quantum here, due to no-cloning or rather no-broadcasting. 

\section{Discussion
%Conclusion
}\label{sec3}

%\section{Possible information volatility beyond teleportation} \label{dilip-ghosh}

%\textbf{\emph{In the distribution of entangled state between parties:}}

The  instances of volatility of information in the preceding sections were always related to teleportation protocols. We believe however that one can find occurrences of information volatility in other protocols as well. 

% Before we conclude, we discuss another scenarios where information appears to remain in volatile form. 
Consider, e.g., a scenario, where an observer, Chippu~\cite{chaTa-high-school}, prepares the 
%Debu that some observer Charu prepared the following state 
two-qubit state, 
\begin{equation}
|\Psi\ket=\frac{1}{\sqrt{2}}\left( |\psi\psi\ket+|\psi^\perp\psi^\perp\ket\right),
\end{equation}
where $|\psi\ket=\cos\frac{\theta}{2}|0\ket+e^{i\phi}\sin\frac{\theta}{2}|1\ket$ with $\theta \in [0,~\pi]$ and $\phi \in [0,~2\pi)$, and \(|\psi^\perp\ket = \sin\frac{\theta}{2}|0\ket-e^{i\phi}\cos\frac{\theta}{2}|1\ket\), a state that is orthogonal to $|\psi\ket$.  
Chippu  distributes one part of $|\Psi\ket$ to Alice and the other to Bob. Alice and Bob were told by Chippu about the form of \(|\Psi\rangle\), but 
not about the values of \(\theta\) and \(\phi\). In this situation, Alice and Bob knows that they share a maximally entangled state, but without the knowledge about \(\theta\) and \(\phi\), they cannot use the shared state, say for some quantum communication protocol. 
(One is reminded here of the frequentist-inspired quantum theory of random phenomena~\cite{1946-47}, which deals with a similar situation.) It is plausible that this situation can give rise to a situation where information about \(|\psi\rangle\) can become volatile, especially in the many-copy situation.

 Researchers 
 have been talking about quantum information already for decades, but it is still far from being explored thoroughly. 
 Undoubtedly, the view that ``information is physical'' played a pivotal role in  classical and quantum information science, in particular on reversible computation and quantum computation. 
 %This notion gave rise to physics of computation
% which has a deep impact on the development of reversible computation and quantum computation. 
It is interesting therefore that information, both quantum and classical, can remain in volatile form without a physical body. We found that the phenomenon is natural to the teleportation protocol, and that the volatility is independent of whether or not we assume special relativity. We believe that the observation can provide important insights  %Nevertheless, it is quite remarkable that  
 %quantum information could exists without a physical body. The possibility of existence of information without physical body we call it as volatileness of quantum information. 
% We have defined the volatileness of quantum information and have shown that this naturally occurs in quantum 
% teleportation. We have also argued that even if we do not assume special relativity principles, quantum information can still remain volatile. We have provided other possible 
% scenario where volatileness of quantum information can happen. We believe that our observation can bring deep insights 
into the nature of information and its connection to 
 physical bodies. It will be interesting to understand the connection of the findings to the problem of the Maxwell's demon. And following the steps of J.A. Wheeler~\cite{mone-poRe-ruby-ray}, we can say that \emph{bit and qubit can remain without it}. 

\acknowledgements
We acknowledge support from the Department of Science and Technology, Government of India through the QuEST grant
(grant numbers  DST/ICPS/QUST/Theme-1/2019/117 and DST/ICPS/QUST/Theme-3/2019/120).

\end{document}